# The instrumentation program at the Large Binocular Telescope Observatory in 2024

Joseph C. Shields*[a], Jason Chu[a], Albert Conrad[a], Jonathan Crass[b,c], Justin R. Crepp[c], Steve Ertel[d], Jacopo Farinato[e], Ilya Ilyin[f], Olga Kuhn[a], Luca Marafatto[e], Fernando Pedichini[g], Roberto Piazzesi[g], Richard W. Pogge[c], Jennifer Power[a], Sam Ragland[a], Robert Reynolds[a], James Riedl[a], Mark Smithwright[a], Klaus G. Strassmeier[f], David Thompson[a]

[a]Large Binocular Telescope Observatory, 933 N. Cherry Ave., Tucson, AZ, USA 85721;
[b]Department of Astronomy, The Ohio State University, Columbus, OH, USA 43210;
[c]Department of Physics & Astronomy, University of Notre Dame, Notre Dame, IN, USA 46556;
[d]Steward Observatory, University of Arizona, Tucson, AZ 85721;
[e]INAF Osservatorio Astronomico di Padova, 35122 Padova, Italy;
[f]Leibniz-Institute for Astrophysics (AIP), 14482 Potsdam, Germany;
[g]INAF Osservatorio Astronomico di Roma, 00078 Monte Porzio Catone, Italy

## ABSTRACT

The Large Binocular Telescope, with its expansive collecting area, angular resolving power, and advanced optical design, provides a robust platform for development and operation of advanced instrumentation for astronomical research. The LBT currently hosts a mature suite of instruments for spectroscopy and imaging at optical through mid-infrared wavelengths, supported by sophisticated adaptive optics systems. This contribution summarizes the current state of instrumentation, including upgrades to existing instruments and commissioning of second generation instruments now in progress. The LBT is soliciting proposals for next generation instrument concepts, with participation open to consortium members and others interested in participation in the Observatory.

## 1. INTRODUCTION

The year 2024 marks the 20th anniversary of the dedication of the Large Binocular Telescope, providing an opportunity to reflect on its history and future prospects, as enabled by the technology embodied in the telescope and its instrumentation. The LBT remains distinctive as the first of the Extremely Large Telescopes, based on multi-mirror technology providing the light-collecting power equivalent to a single-aperture 11.8m telescope, and resolving power provided by a 22.8m edge-to-edge dual aperture.

At the present time the LBT hosts a mature suite of instruments as listed in Table 1, providing versatile capability for optical and infrared imaging and spectroscopy in a variety of configurations. The instrument complement reflects choices intended to leverage specific strengths and attributes of the telescope. The latter properties include i) large collecting area and interferometric capability; ii) a facility-type, distributed, single conjugate adaptive optics system, based on use of adaptive secondary mirrors and pyramid wavefront sensors; iii) use of swingarms above the primary mirrors, enabling rapid reconfiguration between instruments or optical fiber feeds located at prime focus, Gregorian focus, or bent Gregorian focus; and iv) the Observatory site at Mt Graham, which at 3200m is among the highest altitude observatory sites in North America, with corresponding reduction in absorption by atmospheric water vapor.

The first-generation instruments at the LBT have been described in multiple papers in the literature, and reviewed most recently by Rothberg et al. (2020)[1]. In the present work we describe recent activity resulting in changes in existing instrumentation, and an overview of second-generation instruments in various stages of development and commissioning.

*jshields@lbto.org

Recent developments in adaptive optics at LBTO are addressed in other papers in this conference (Ragland et al. 2024[2], Brusa et al. 2024[3], Guerra et al. 2024[4], Veillet et al. 2024[5], Zhang et al. 2024[6]).

Table 1. Current instruments available for use at the LBT. Instruments fall into three classes: Facility Instruments funded by the LBT consortium as a whole, and supported by the Observatory for general use by consortium Member astronomers; Strategic Instruments, designed for interferometric capability with substantial funding and operational support from a subset of Members; and Principal Investigator (PI) instruments, which are developed and funded by one or more Members, who retain control and responsibility for the instrument.

| | |
|---|---|
| Facility Instruments | |
| Large Binocular Cameras (LBCs) | 0.35 – 1 µm, 23′ FOV |
| Multi-Object Double Spectrographs (MODS) | Imaging + Spectroscopy, 320-1000 nm, 6′×6′ FOV, $R \leq 2000$ |
| LBT Utility Cameras in the Infrared (LUCIs) | Imaging + spectroscopy, 0.95-2.4 µm, 4′ FOV, R = 2000 - 8500 |
| Potsdam Echelle Polarimetric and Spectroscopic Instrument (PEPSI) | 383-912 nm, R=50,000, 130,000, 250,000 and IQUV with R=130,000 |
| L- and M-band Infrared Camera (LMIRCam) | 3-5 µm imaging and IFU with R~100 |
| | |
| Strategic Instruments | |
| LBT Interferometer (LBTI) | Fizeau imaging and nulling, feeds LMIRcam and NOMIC |
| LINC/NIRVANA | Multiconjugate AO system, 2′×2′ FOV (in commissioning) |
| | |
| PI Instruments | |
| Nulling Optimized Mid-Infrared Camera (NOMIC) | 8-13 µm, 9″×9″ FOV |
| SHARK-NIR | 0.96-1.7 µm, 18″×18″ FOV, $R \leq 700$, extreme AO |
| SHARK-VIS | 450-950 nm, 10″diameter FOV, extreme AO |
| iLocater | 970-1310 nm, R=190,000, AO, delivery in 2025 |

## 2. UPDATES TO EXISTING INSTRUMENTATION

**Multi-Object Double Spectrographs**

The Multi-Object Double Spectrographs (MODS; Pogge et al. 2010[7]) are a pair (one for each telescope primary mirror) of high-throughput low- to medium-resolution instruments operating in the 340-1000 nm range. The instruments were built by Ohio State University. The design uses a dichroic to split light into separate red and blue channels, each with 8k×3k E2V CCD detectors operated by OSU MkIX detector controllers. The MODS units were installed at the LBT in 2010 and 2014, and have proven to be highly reliable and productive, generating data for more publications than any other instruments at the Observatory.

One emerging concern in recent years is that the ability to maintain the MODS CCD controllers has diminished due to changes in personnel and technology, and no spares are available for these systems, giving rise to a serious risk of single-point failure. The Observatory consequently initiated a program to replace the controllers and associated DOS computers with modern elements that are more readily supported and maintained. In 2023 six STA Archon controllers were purchased, sufficient for the 4 CCDs with 2 spares. Development of interface electronics as well as software is currently in progress, led by Ohio State with participation also by the University of Arizona. Deployment of the four new systems in the two MODS instruments is planned for completion during the 2025 summer shutdown.

**Potsdam Echelle and Polarimetric Spectrograph**

The Potsdam Echelle Polarimetric and Spectroscopic Instrument (PEPSI; Strassmeier et al. 2015[8], 2018[9]) delivers spectra spanning 383-912 nm with resolution up to 250,000. The instrument is mounted on a bench in a stabilized chamber installed on the ground floor in the telescope pier, fed by optical fibers at the bent Gregorian foci. The instrument separates light into red and blue arms via a dichroic, with three cross dispersers on each side enabling coverage of the full spectral range in three exposures, recorded on each arm with a STA 1600LN 10.3k×10.3k CCD. PEPSI provides spectropolarimetric capability via fiber feeds at the Gregorian focus, and additionally provides daytime spectropolarimetric Stokes IQUV solar monitoring via a robotic Solar Disk Integration unit. PEPSI can also be fed by the 1.8m Vatican Advanced Technology Telescope on Mt. Graham via a 400m fiber link (Strassmeier et al. 2023[10]). PEPSI was initiated as a PI instrument, and officially accepted as a Facility Instrument in 2015, with personnel from AIP-Potsdam continuing to provide critical technical support. PEPSI has proven to be a powerful instrument for stellar astrophysics and exoplanet transit studies, and is popular also as a backup instrument for use in marginal observing conditions.

PEPSI has undergone several minor upgrades in recent years, with a major transition occurring via replacement of its detectors. The original CCDs displayed a flux-dependent fixed pattern noise as well as amplifier glow, particularly on the red side. To correct these issues, new STA CCDs were installed during the 2023 summer shutdown, yielding much improved noise characteristics.

**LBT Interferometer**

The LBT Interferometer (LBTI; Hinz et al. 2016[11]; Ertel et al. 2020[12]) operates at the bent Gregorian focus as a beam combiner, designed to feed other instruments. Light enters LBTI after encountering only three warm optical components (M1, M2, and M3 mirrors), and is combined from the two sides incoherently, or coherently with either Fizeau imaging or a nulling mode. The LMIRCam and NOMIC instruments are operated in conjunction with LBTI and use the telescope's AO capability. The increased availability and reliability of the AO system has translated directly into increased LBTI observations, with results to date focusing mostly on circumstellar disks, exoplanets, and related phenomena (e.g., Miles et al. 2024[13]; Isbell & Ertel 2024[14]).

An essential part of LBTI for interferometry is the fringe tracker based on PHASECam (Defrère et al. 2014[15]), an imager that uses a Rockwell PICNIC array. This detector is over 20 years old, and plans are underway to replace it with the Fizeau Fringe Tracking Camera (FFTCam; Conrad et al. 2023[16]) which makes use of a modern Saphira avalanche photodiode detector. Development is being led by the U.S. Naval Research Laboratory and the University of Arizona, with installation planned for the second half of 2024. The impact on observing capability is expected to be profound, with an increase in limiting magnitude for reference stars from the current K=4.7 to K=10. The 100× increase in depth will expand the number of potential reference stars from ~$5 \times 10^4$ to ~$10^7$.

## 3. SECOND GENERATION INSTRUMENTS

**SHARKs**

The System for coronagraphy with High order Adaptive optics from R to K band (SHARK) is implemented with two channels, SHARK-NIR (near-infrared, Farinato et al. 2022[17]) and SHARK-VIS (visible, Pedichini et al. 2022[18], 2024[19]). The SHARK instruments are optimized for high contrast, high spatial resolution measurements, with a primary science focus on exoplanet studies. The instruments have compact designs that allow them to share focal stations with LBTI, with SHARK-NIR installed on the left (SX) side of the telescope, and SHARK-VIS installed on the right (DX) side. The capability of observing the same target simultaneously with SHARK-VIS, SHARK-NIR, and LMIRCam at wavelengths from the visible to the thermal infrared has been demonstrated and is expected to become routine in fall 2024. Any combination of two of these instruments is already routinely used simultaneously.

**SHARK-NIR**

SHARK-NIR provides a camera for high-contrast imaging and spectroscopy between 0.96-1.7 μm, with observing modes available for classical imaging, coronagraphic imaging, dual band imaging, and long slit spectroscopy (R=100 and 700). Development has been led by INAF-Padova with participation also by the University of Arizona, Max Planck Institute for Astronomy, and the Institut de Planétologie et d'Astrophysique de Grenoble.

SHARK-NIR was installed and commissioning began in 2022, with science observations initiated in 2023. Initial results are very promising, with coronagraphy yielding contrasts of $\sim 10^{-4} - 10^{-5}$ at separations of 0.″1 – 0.″3 for stars with H-band magnitude of 5-6 (Barbato et al. 2024[20]).

**SHARK-VIS**

SHARK-VIS provides a camera for high-contrast imaging in the 400-1000 nm interval, within which Hα is of specific interest as a tracer of accreting young exoplanets. Development has been led by INAF-Rome with participation also by the University of Arizona.

SHARK-VIS was installed in 2023 with commissioning and science observations following soon after (Pedichini et al. 2024[19]). The first peer-reviewed results (Conrad et al. 2024[21]) present multi-bandpass imaging of Jupiter's moon Io. The data achieve a spatial resolution of ~24 mas, corresponding to ~80 km on Io's surface. The measurements reveal changes in surface features linked to volcanic ejecta most likely generated within the past 4 years. The detail visible in the SHARK-VIS images exceeds that in Hubble Space Telescope observations of Io, due to the larger primary mirror at the LBT, and demonstrate a new regime for planetary imaging from the ground.

**iLocater**

The iLocater instrument (Crepp et al. 2016[22], Crass et al. 2024[23]) is a high resolution spectrograph operating at 970-1310 nm wavelength with median R=190,000, optimized for extreme precision radial velocity measurements of stars for study of associated exoplanets. The cryogenic instrument uses single-mode fibers with diffraction-limited optics allowing for a compact design. iLocater and its associated cryostat are planned for installation in 2025 in an environmentally controlled chamber on Level 3 in the telescope pier. The fiber injection system will share a bent-Gregorian focal station with LBTI. Development of iLocater is led by the University of Notre Dame and Ohio State University.

## 4. FUTURE OPPORTUNITIES

The Large Binocular Telescope Observatory is currently inviting inquiries from additional institutions that may have interest in participating in this forefront research facility. The Observatory is operated by the University of Arizona under a contract with the LBT Corporation (LBTC), a collaboration of leading scientific institutions in the U.S. and Europe. As outlined in this summary, the LBT offers rich opportunities for advancing science with its existing instrument complement, while offering an attractive platform for new instrument concepts and development.

Instrumentation at the LBT has been driven from the outset by the interests of researchers at its member institutions. The telescope has a demonstrated track record of innovation in instrumentation and telescope technology, and is in a strong position to support further innovation now and in the future. With second-generation instruments now achieving fruition, the LBTC is issuing a call for proposals for new instrument concepts, with an initial review scheduled for spring 2025. The call is open to LBT Members, as well as other institutions that may wish to pursue instrument development in conjunction with participation in the Observatory. Collaboration between LBT Members and researchers beyond the LBT consortium in the development of new instrument concepts is encouraged.

## ACKNOWLEDGEMENTS

The LBT is an international collaboration among institutions in the United States, Italy, and Germany. LBT Corporation Members are: The University of Arizona on behalf of the Arizona Board of Regents; Istituto Nazionale di Astrofisica, Italy; LBT Beteiligungsgesellschaft, Germany, representing the Max Planck Society, the Leibniz Institute for Astrophysics Potsdam, and Heidelberg University; and The Ohio State University, representing OSU, University of Notre Dame, University of Minnesota, and University of Virginia.

## REFERENCES

[1] Rothberg, B., et al., "Current status of the facility instruments at the Large Binocular Telescope Observatory," Proc. SPIE 11447, Ground-based and Airborne Instrumentation for Astronomy VIII, 1144706 (2020). https://doi.org/10.1117/12.2563352

[2] Ragland, S., et al., "Emerging adaptive optics facility at the Large Binocular Telescope," Proc. SPIE 13097, Adaptive Optics Systems IX, 13097-20 (2024).

[3] Brusa, G., et al., "Adaptive secondary mirrors at the Large Binocular Telescope recent updates," Proc. SPIE 13097, Adaptive Optics Systems IX, 13097-95 (2024).

[4] Guerra, J. C., et al., "On-sky performance metric parametrization@LBTO pyramid WFS," Proc. SPIE 13097, Adaptive Optics Systems IX, 13097-243 (2024).

[5] Veillet, C., Guerra, J. C., Masciadri, E., and Ragland, S., "Turbulence monitoring at the Large Binocular Telescope: current status and ongoing developments," Proc. SPIE 13097, Adaptive Optics Systems IX, 13097-310 (2024).

[6] Zhang, X., et al., "AO functionalities improvements with the adaptive secondary mirrors at the Large Binocular Telescope," Proc. SPIE 13097, Adaptive Optics Systems IX, 13097-264 (2024).

[7] Pogge, R. W., et al., "The multi-object double spectrographs for the Large Binocular Telescope," Proc. SPIE 7735, Ground-based and Airborne Instrumentation for Astronomy III, 77350A (2010). https://doi.org/10.1117/12.857215

[8] Strassmeier, K. G., et al., "PEPSI: The high-resolution echelle spectrograph and polarimeter for the Large Binocular Telescope," Astron. Nachr., 336, 324 (2015). https://doi.org/10.1002/asna.201512172

[9] Strassmeier, K. G., et al., "Want a PEPSI? Performance status of the recently commissioned high-resolution spectrograph and polarimeter for the 2x8.4m Large Binocular Telescope," Proc. SPIE 10702, Ground-based and Airborne Instrumentation for Astronomy VII, 1070212 (2018). https://doi.org/10.1117/12.2311627

[10] Strassmeier, K. G., et al., "VPNEP: Detailed characterization of TESS targets around the Northern Ecliptics Pole. I. Survey design, pilot analysis, and initial data release," Astron. & Astrophys., 671, A7 (2023). https://doi.org/10.1051/0004-6361/202245255

[11] Hinz, P. M., et al., "Overview of LBTI: a multipurpose facility for high spatial resolution observations," Proc. SPIE 9907, Optical and Infrared Interferometry and Imaging V, 990704 (2016). https://doi.org/10.1117/12.2233795

[12] Ertel, S., et al., "Overview and prospects of the LBTI beyond the completed HOSTS survey," Proc. SPIE 11446, Optical and Infrared Interferometry and Imaging VII, 1144607 (2020). https://doi.org/10.1117/12.2561849

[13] Miles, B. E., et al., "Sensitivity and performance of LBTI/NOMIC spectroscopy: prospects for 10- and 30-meter class mid-IR exoplanet science," Proc. SPIE 13096, Ground-based and Airborne Instrumentation for Astronomy X, 13096-224 (2024).

[14] Isbell, J., and Ertel, S., "The LBTI: pioneering the ELT era," Proc. SPIE 13095, Optical and Infrared Interferometry and Imaging IX, 13095-5 (2024).

[15] Defrère, D., et al., "Co-phasing the Large Binocular Telescope: status and performance of LBTI/PHASECam," Proc. SPIE 9146, Orbital and Infrared Interferometry IV, 914609 (2014). https://doi.org/10.1117/12.2057178

[16] Conrad, A., Stone, J., and Ertel, S., "Increased sky coverage for the 23-meter LBT," Adaptive Optics for Extremely Large Telescopes (AO4ELT7) (2023). https://dx.doi.org/10.13009/AO4ELT7-2023-039

[17] Farinato, J., et al., "SHARK-NIR, ready to 'swim' in the LBT Northern Hemisphere 'ocean'," Proc. SPIE 12185, Adaptive Optics Systems VIII, 1218522 (2022). https://doi.org/10.1117/12.2630083

[18] Pedichini, F., et al., "SHARK-VIS ready for the stars: instrument description and final laboratory test," Proc. SPIE 12185, Adaptive Optics Systems VIII, 121856Q (2022). https://doi.org/10.1117/12.2629244

[19] Pedichini, F., et al., "AO assisted visible observations with the new high contrast imager SHARK-VIS at the Large Binocular Telescope: early results," Proc. SPIE 13097, Adaptive Optics Systems IX, 13097-9 (2024).

[20] Barbato, D., et al., "SHARK-NIR commissioning and early science runs," Proc. SPIE 13096, Ground-based and Airborne Instrumentation for Astronomy X, 13096-68 (2024).
[21] Conrad, A., et al., "Observation of Io's Resurfacing via Plume Deposition Using Ground-Based Adaptive Optics at Visible Wavelengths with LBT SHARK-VIS," Geophys. Res. Lett., 51, e2024GL108609 (2024). https://doi.org/10.1029/2024GL108609
[22] Crepp, J., et al. 2016, "iLocater: a diffraction-limited Doppler spectrometer for the Large Binocular Telescope," Proc. SPIE 9908, Ground-based and Airborne Instrumentation for Astronomy VI, 990819 (2016). https://doi.org/10.1117/12.2233135
[23] Crass, J, et al. 2024, "Performance characterization of the iLocater spectrograph," Proc. SPIE 13096, Ground-based and Airborne Instrumentation for Astronomy X, 13096-50 (2024).